\documentclass{osa-article}

\journal{osac}

\articletype{Research Article}

\usepackage{float}
\usepackage{amsmath}
\usepackage{xcolor}
\usepackage{tabto}

\begin{document}

\title{Offset Lock with 440 GHz Range using Electro-Optic Modulation}

\author{Ocean Zhou \authormark{1, 3, *, $\dagger$}, Andrew O. Neely \authormark{1,4, *}, Zachary R. Pagel \authormark{1, 2}, Madeline Bernstein \authormark{1, 2}, Jack Roth \authormark{1}, and Holger Mueller \authormark{1, 2, $\dagger\dagger$}}

\address{\authormark{1} Physics Department, University of California, Berkeley, South Hall Rd, Berkeley, CA 94720, USA\\
\authormark{2} Lawrence Berkeley National Laboratory, One Cyclotron Road, Berkeley, California 94720, USA\\
\authormark{3} Now at Applied Physics Department, Stanford University, 348 Via Pueblo, Stanford, CA 94305, USA\\
\authormark{4} Now at Physics Department, Yale University, 217 Prospect St, New Haven, CT 06511, USA \\
\authormark{*} Both individuals contributed equally to this paper.}

\email{\authormark{$\dagger$}ozhou619@stanford.edu}
\email{\authormark{$\dagger\dagger$}hm@berkeley.edu}
 
\begin{abstract}
 Offset locking is crucial to many physics experiments. Wide range offset locks are desirable, as they increase the span of usable frequencies in an experiment. Here, we experimentally realize a wide-range offset lock using a beat-note setup combined with electro-optic phase modulation. By using frequency down-conversion of the beat note and locking to sidebands generated by electro-optic modulation, we achieve an offset range of $\pm$ 220.1 GHz with offset frequency fluctuations under 0.1 Hz and a phase error variance of 0.017 rad$^2$ over a 100 kHz bandwidth, greatly widening the range compared to past setups using this method. The relative simplicity of our setup provides a compelling method for locking at offsets in the hundreds of GHz range.
\end{abstract}

\section{Introduction}

\begin{figure}[!b]
    \centering
    \includegraphics[width = \textwidth]{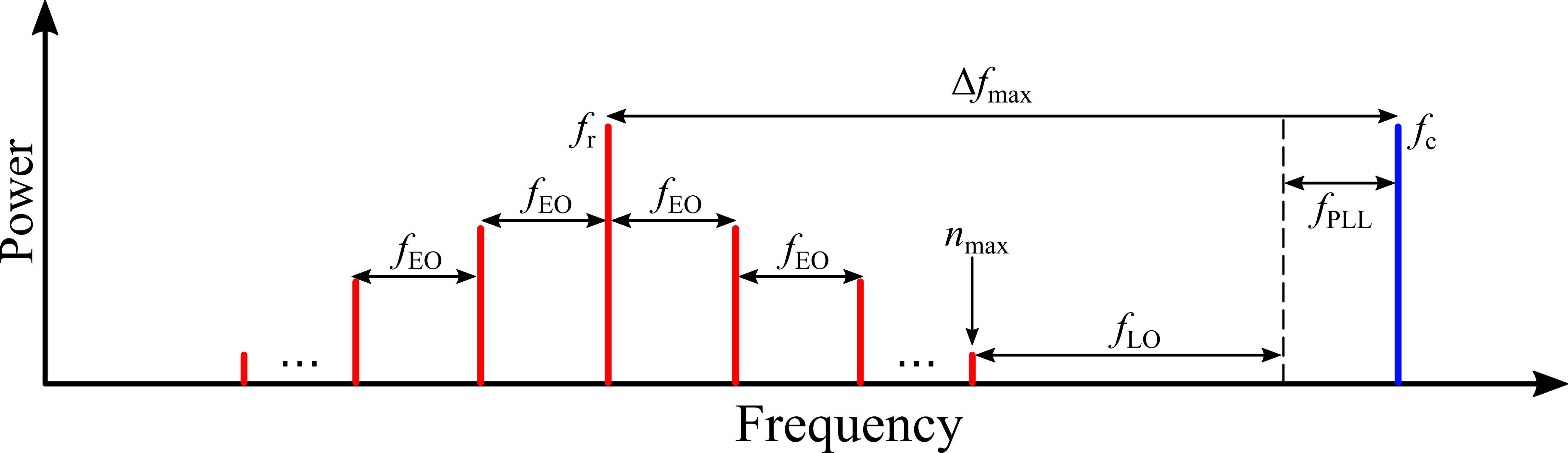}
    \caption{Diagram showing the maximum offset frequency of our setup. The reference laser frequency is $f_\mathrm{r}$ and the controlled laser frequency is $f_\mathrm{c}$. The EOM is modulated at frequency $f_\mathrm{EO}$ while the local oscillator (LO) frequency used for down-conversion is $f_\mathrm{LO}$. The frequency of the PLL reference input is $f_\mathrm{PLL}$. The sidebands are represented by the red lines on both sides of $f_\mathrm{r}$. The maximum sideband order used is $n_{\mathrm{max}}$. To achieve the maximum offset, we use $f_{\mathrm{EO}} = 20$ GHz, $f_{\mathrm{LO}} = 40$ GHz, $n_{\mathrm{max}}$ = 9, and $f_{\mathrm{PLL}} = 100$ MHz.}
    \label{fig: freqdiagram}
\end{figure}%

\tab Stabilizing ("locking") laser frequency is crucial to a wide range of applications, including metrology \cite{metrology}, greenhouse gas spectroscopy \cite{spectroscopy}, and quantum computation \cite{QMComp}. Many experiments use offset locking to stabilize laser frequency, with examples including laser spectroscopy \cite{3GHzOffsetWater}, laser cooling \cite{46GHzOffset, 10GHzSimpleOffset, SideOfFilterLock} and imaging of ultracold atoms \cite{Offsetimaging}. Offset locking stabilizes the frequency of a controlled laser against a reference laser by locking the frequency difference between the two \cite{10GHzSimpleOffset, 20GHzLock, 46GHzOffset, 5THzOffset, SideOfFilterLock, Offsetdelayline, OffsetDFB, OffsetEIT, OffsetFMS, OffsetFtoV, Offsetimaging, OffsetNdYag, 3GHzOffsetWater, Rev1OptLett}. An offset lock can be an optical phase locked loop (OPLL) to ensure that the controlled laser is coherent with the reference laser, which decreases fluctuations in laser frequency \cite{OPLL1st, OPLLDiode, OPLLF106mu, OPLLF9GHzDiode, OPLLF9GHzECSL, OPLLIntHet, OPLLperf, OPLLsemicond, OPLLSingleSide, PLLDiodeLaser, OffsetEOM, OPLLDFB, OPLLECDL}.

While many methods exist for achieving wide locking ranges, such methods often require more complicated and expensive components \cite{5THzOffset, Rev1IEEE}. Transfer cavity locks can possess ranges in the few-THz but require intricate optical construction and careful alignment \cite{5THzOffset, TF2, TF3, TF4}. Very large ranges have also been obtained by optical frequency combs generated by using strong, intracavity phase modulation with low-loss electro-optic modulators (EOMs) \cite{Rev1IEEE}. However, intracavity EOM locks have high temperature sensitivity and are coupled to greater degrees of freedom \cite{Rev1IEEE}. Offset locks employing Mach-Zehnder modulators (MZMs) can also reach ranges between hundreds of GHz up to THz \cite{Rev2RollandOptExp, Rev2RollandIEEE, Rev2DelHayePRL, Rev2JiangLiSci}. We present a simpler alternative, based on previous methods \cite{46GHzOffset, OffsetEOM}, that has a range in the hundreds of GHz and relies only on an optical heterodyne beat note combined with electro-optic modulation of one of the two beams, frequency down-conversion, and a PLL. We phase modulate the reference laser with an electro-optic modulator (EOM), creating higher order frequency sidebands, as shown in Fig. \ref{fig: freqdiagram}. Then, we create a beat note between the controlled laser beam and the modulated reference beam. The beat note is frequency down-converted using a local oscillator (LO) signal to extend the range and to also ensure that the beat note frequency is within the bandwidth of the subsequent electronics. This is input to a PLL, generating a control signal that adjusts the phase of the controlled laser so that the beat note is coherent with a PLL reference input signal \cite{PLLBook1, PLLBook2, PLLDiodeLaser}. Ensuring phase coherence between the two signals means the PLL locks the down-converted beat note frequency so that it is equal to the PLL reference frequency, which consequently stabilizes the controlled laser frequency. We further extend the lock range by locking with higher sideband orders of the beat note, as shown in Fig. \ref{fig: freqdiagram}, and achieve an offset range of $\Delta f_{\mathrm{max}} = \pm$ 220.1 GHz, which widens the range up to 10 times compared to previous efforts using this method \cite{46GHzOffset, OffsetEOM}.

\section{Experimental Setup \label{sec: exp}}



\begin{figure}[!t]
    \centering
    \includegraphics[width = \textwidth]{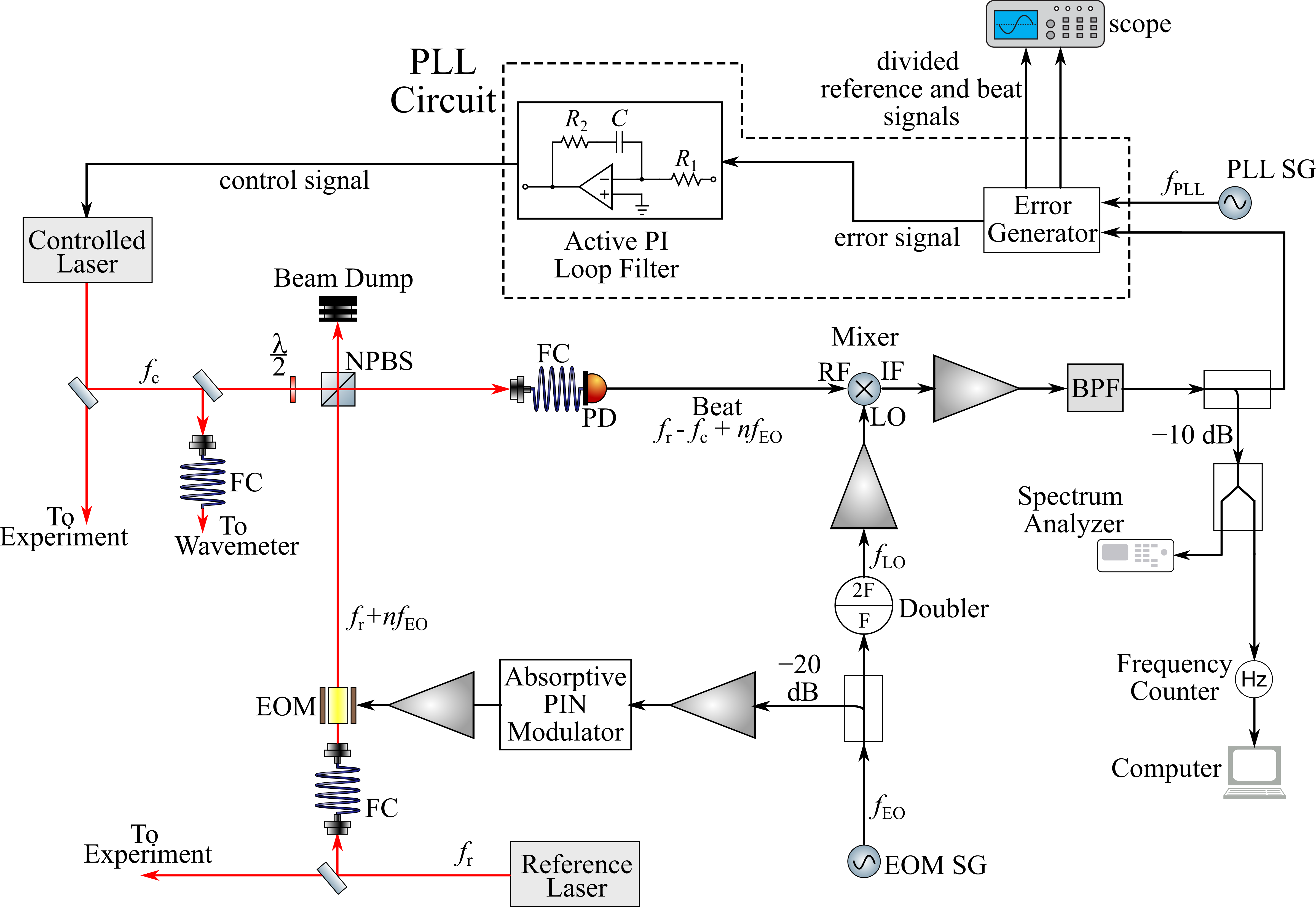}
    \caption{Layout of the wide range EOM offset lock. Key: NPBS, non-polarizing beam splitter; BPF, bandpass filter; EOM, electro-optical modulator; BS, beam splitter; PD, photodiode; SG, signal generator; FC, Fiber Cable; RF, radio-frequency; PI, proportional-integral; $\frac{\lambda}{2}$, half-wave plate.}
    \label{fig: expsetup}
\end{figure}

\tab Our offset lock setup is shown in Fig. \ref{fig: expsetup}. The controlled laser is a Ti:Sapphire ring laser \cite{M2Laser}. Along the controlled laser arm, part of the light is sent into a fiber connected to a wavemeter, used to monitor the laser frequency. The reference laser is an external cavity diode laser that is spectroscopically locked to an atomic transition with wavelength approximately at 850 nm \cite{ECDL}. Details on the specific parts of the lock can be found in Section 1 of the supplemental document.

The reference beam is phase-modulated by a fiberized lithium niobate EOM that has a bandwidth of 20 GHz and a half-wave voltage of 2.2 V. The two beams are superimposed using a non-polarizing beam splitter (NPBS) onto a photodiode with 40 GHz RF bandwidth to create a beat note. The half-wave plate before the NPBS is used to ensure the two beams have the same polarization. The beat is down-converted with a LO signal created from a microwave signal generator (SG), labeled EOM SG in Fig. \ref{fig: expsetup}. The EOM SG signal is input to a -20 dB coupler. The coupled signal is amplified, adjustably attenuated by an absorptive PIN modulator and then amplified again to drive the EOM. We use the absorptive PIN modulator to control the amount of drive power delivered to the EOM by adjusting the amount of attenuation on the amplified EOM SG signal. The rest of the EOM SG signal after the coupler is frequency doubled, amplified and sent to the mixer as the LO signal for down-conversion. To achieve the maximum offset frequency, we drive the EOM at frequency $f_{\mathrm{EO}} = 20$ GHz and down-convert with a LO frequency of $f_{\mathrm{LO}} = 2f_{\mathrm{EO}}$ = 40 GHz.

The down-converted beat signal is amplified and bandpass-filtered. The beat signal is input to a -10 dB coupler, with the coupled signal then sent to a power splitter, directing the beat signal into a spectrum analyzer to measure its power and a frequency counter to measure the down-converted beat frequency. The rest of the beat signal is delivered to the PLL circuit, which consists of an error generator and a proportional-integral (PI) loop filter. The PLL reference signal, also input into the circuit, has a frequency $f_\mathrm{PLL}$ = 100 MHz and is produced by the PLL signal generator, labelled as PLL SG in Fig. \ref{fig: expsetup}. The error generator first converts the two RF inputs into logic signals using a dual comparator \cite{AD96687}. Then, the comparator outputs are each sent to two 4-bit counters and are frequency divided by a factor of $N = 32$ \cite{MC10H016}. The divided signals are sent to a digital phase and frequency discriminator (PFD) which outputs a pulse train with an average value that is proportional to the phase error by a factor $K_d = 0.286$ V/rad \cite{AD9901}. The frequency division is crucial as it enables the PFD to operate in a more linear regime \cite{AD9901}. Finally, the inverting and noninverting PFD outputs are sent to an instrumentation amplifier with unity gain, generating the error signal. The error signal is input to the loop filter, which produces a control signal that is fed back to the controlled laser.


\section{Results \label{sec: res}}

\tab We achieve an offset range of $\Delta f_{\mathrm{max}} = \pm$ 220.1 GHz with $n_{\max} = \pm 9$ and the range being limited by the optical power in the higher order sidebands. We characterize the lock by locking the controlled laser at offset frequencies from $\pm$40.1 GHz to $\pm$220.1 GHz in 20 GHz intervals, corresponding to different EOM sideband orders. At each desired offset frequency, $\Delta f_{\mathrm{des}}$, we adjust the amplitude with which the EOM is driven so that the beat signal power is maximized and then record this maximum power $P_{\mathrm{max}}$ as well as the beat note frequency over a 10 to 20 minute interval. Afterwards, we find the mean offset frequency $\overline{\Delta f}$, the offset variance Var$({\Delta f})$, and the standard deviation in the offset frequency $\sigma_{\mathrm{\Delta} f}$. The offset variance is defined as Var$({\Delta f}) = (\overline{\Delta f} - \Delta f_{\mathrm{des}})^2$ and is not a measure of actual offset frequency error but rather the noise about the offset frequency. The results are plotted in Fig. \ref{fig: results}(a) and (b) and tabulated in the supplemental document as Table S1. We observe that the effective range of the lock spans $\Delta f_{\mathrm{max}} = \pm$ 220.1 GHz. Within this range, the beat signal is strong enough for the PLL to successfully lock, leading to lower phase noise. Accordingly, we observe $\sigma_{\Delta f}$ to be around 60 mHz and Var$({\mathrm{\Delta} f})$ to be under $10^{-4}$ Hz$^2$. Outside of this range, a phase lock is no longer present as the beat signal power dips below a critical threshold. Consequently, the phase noise can no longer be sufficiently attenuated, with $\sigma_{\Delta f}$ and Var$({\mathrm{\Delta} f})$ increasing drastically. Additional frequency stability analysis is shown in Section 2 of the supplemental document.

\begin{figure}[H]
    \centering
    \includegraphics[width = \textwidth]{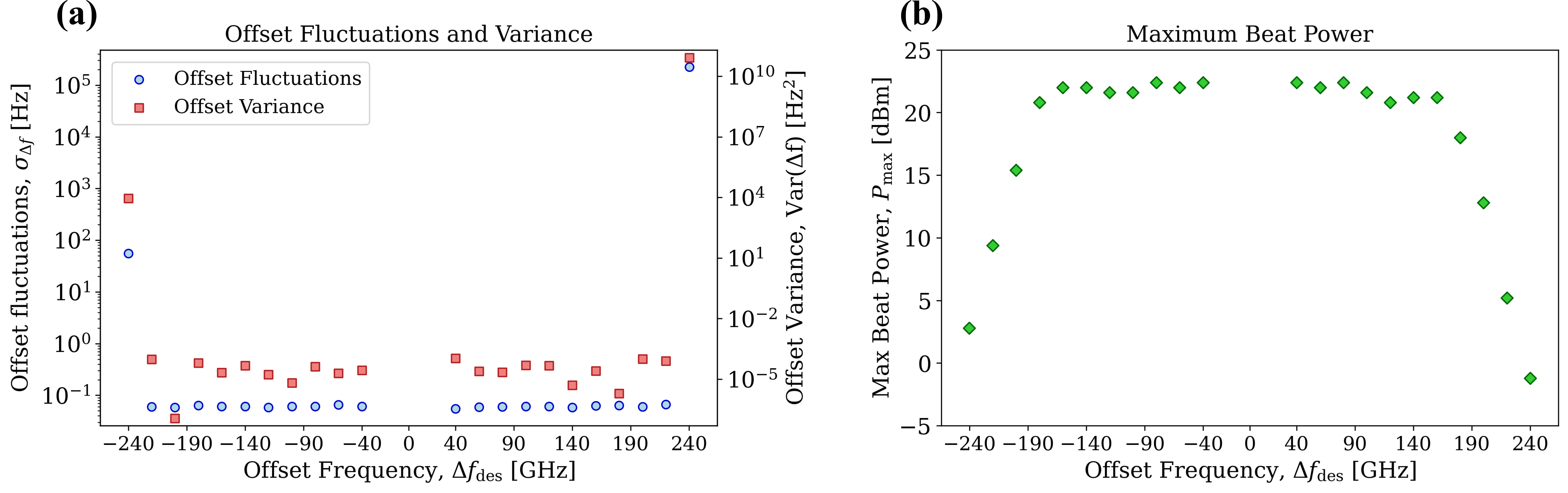}
    \caption{\textbf{(a).} Offset frequency fluctuations, $\sigma_{\mathrm{\Delta} f}$, and offset variance, Var(${\mathrm{\Delta} f}$), versus offset frequency, $\Delta f_{\mathrm{des}}$. For offsets within the range $\Delta f_{\mathrm{max}} = \pm$ 220.1 GHz, the small $\sigma_{\mathrm{\Delta} f}$ and Var$({\mathrm{\Delta} f})$ indicate we have a phase lock. At offsets of $\pm$ 240.1 GHz, we observe large increases in $\sigma_{\mathrm{\Delta} f}$ and Var$({\mathrm{\Delta} f})$, indicating we no longer have a phase lock. The range of the frequency counter is 1 mHz to 400 MHz \textbf{(b).} Maximum beat signal power, $P_\mathrm{max}$, versus $\Delta f_{\mathrm{des}}$. $P_{\mathrm{max}}$ hovers around 21 dBm for offsets between $+$160.1 GHz and -180.1 GHz and decreases exponentially outside of this range.}
    \label{fig: results}
\end{figure}

\begin{figure}[H]
    \centering
    \includegraphics[width = \textwidth]{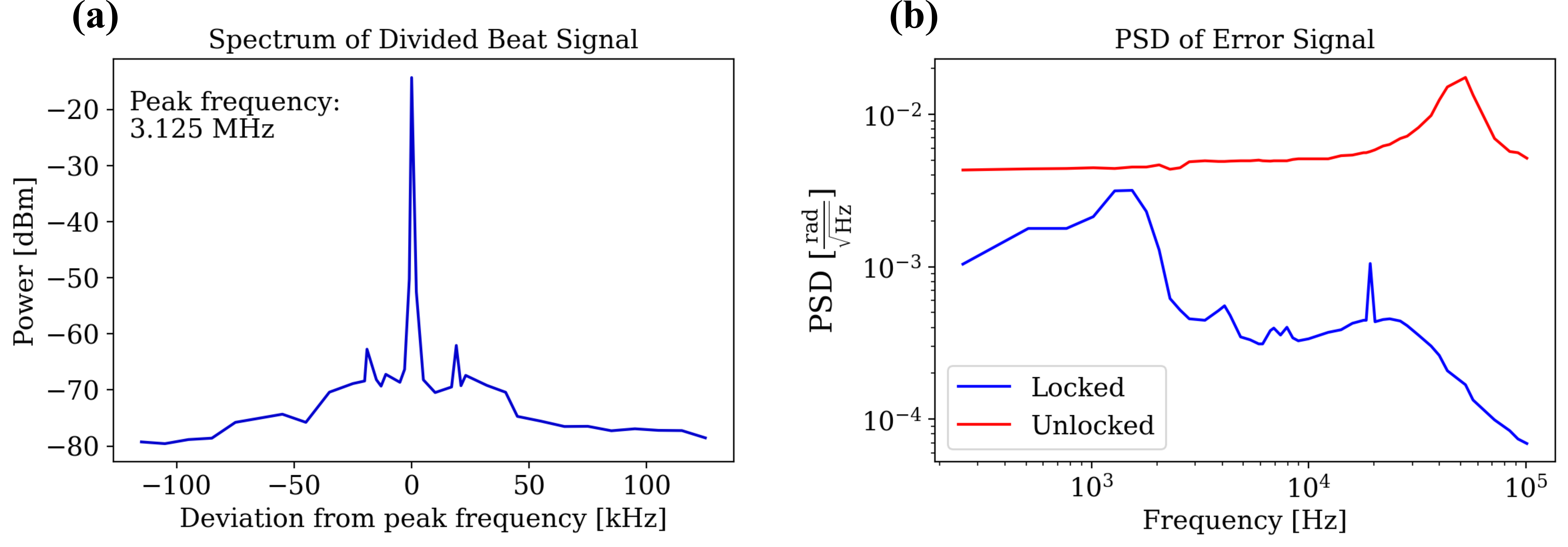}
    \caption{\textbf{(a).} Recorded single-shot spectrum of the divided beat note signal. Resolution bandwidth of the spectrum analyzer was 100 Hz. \textbf{(b).} Power spectral density (PSD) of the error signal when the phase is unlocked versus locked, which shows that the loop filter suppresses phase noise when the controlled laser is locked and the PLL is closed. The phase error variance is calculated to be 0.017 rad$^2$ over a bandwidth of 100 kHz by integrating the error signal PSD.}
    \label{fig: PLLresults}
\end{figure}

We confirm a successful PLL by viewing the divided PLL reference signal and divided beat signal on a scope and observing coherence. Our beat note spectrum matches the expected shape found in previous studies \cite{OPLLFPGA, PLLDiodeLaser, OPLLDiode, OPLLF9GHzDiode, OPLLF9GHzECSL}, where we observe a sharp peak at $3.125$ MHz representing the locked beat signal. The two peaks on the side at $\pm$ 19 kHz away from the central peak are due to pickup from the power supplies. When the phase lock is opened, the central peak of the beat signal spectrum in \ref{fig: PLLresults}(a) becomes less sharp and the surrounding noise floor becomes flat. 

Finally, we record the power spectral density (PSD) of the phase noise shown in Fig. \ref{fig: PLLresults}(b), which demonstrates the loop filter's attenuation of the phase noise across the entire spectrum of measurement when the lock is closed. The narrow peak at 19 kHz in the "locked" curve corresponds to the pickup observed in the beat note spectrum. The broad peak around 24 kHz in the "locked" curve is a servo bump, indicating the measured loop bandwidth \cite{OPLLFPGA}. We calculate  the phase error variance to be 0.017 rad$^2$ over a 100 kHz bandwidth. 

\section{Conclusion \label{sec: con}}

\tab We achieve a wide range offset lock by utilizing electro-optic modulation combined with frequency down-conversion. We measure the range to be 440.2 GHz wide with offset fluctuations under 0.1 Hz and a phase error variance of 0.017 rad$^2$ over a 100 kHz bandwidth. This range is a significant improvement over past studies that employ the same method. The simplicity of our setup, which involves no optical cavity nor custom components, makes our method attractive for locking at offsets in the hundreds of GHz range compared to other methods that have achieved similar or higher locking ranges \cite{5THzOffset, Rev1IEEE, Rev1OptLett, Rev2DelHayePRL, Rev2RollandIEEE, Rev2RollandOptExp}. As such, our work widens the applicability of optical heterodyne offset locking by increasing its range while maintaining a simple cost-effective setup. Possible improvements include using two EOMs to modulate both lasers, and employing a higher bandwidth EOM to increase $f_{\mathrm{EO}}$, which would both further widen the range of the lock.

\begin{backmatter}
\bmsection{Funding}
W. M. Keck Foundation (042982); National Science Foundation (NSF) (1806583); Department of Energy (DoE) (7510743)

\bmsection{Acknowledgments}
We thank the funding agencies listed above. We also thank Yair Segev for additional help on reviewing the paper. 

\bmsection{Disclosures}
The authors declare no conflicts of interest.

\bmsection{Data Availability Statement}
The raw data underlying the results presented in this paper are not publicly available at this time but may be obtained from the authors upon reasonable request. 

\bmsection{Supplemental document}
See Supplement 1 for supporting content. 
\end{backmatter}


\bibliography{sample}






\end{document}


\maketitle

\section{Specific Parts Used in Lock}

\tab The controlled laser is an M-Squared SolsTiS-SA-PSX-XF with a linewidth of about 100 kHz \cite{M2Laser} and pumped by a Coherent Verdi V18. The controlled laser's frequency is adjusted by altering the length of the laser's resonator cavity with a voltage sensitivity $K_o$ = 4 MHz/V. The controlled laser is specified to output 5 W and part of this light is picked off and fiber coupled into one arm of the setup. Along the controlled arm, approximately 1 mW is sent to the wavemeter. The reference laser is an AOSense external cavity diode laser (AOS-IF-ECDL-852) with a 75 kHz linewidth \cite{ECDL} and is spectroscopically locked to the $F = 3 \rightarrow F' = 2$ transition of cesium. 6.5 mW of the reference laser light is coupled to the EOM. Due to insertion loss, the reference arm has 4 mW of light after the EOM. After combining both beams, in total, about 8 mW of light hit the photodiode. The EOM is made by EOSpace with part number PM-0S5-20-PFA-PFA-850. The photodiode is made by OptiLab with part number PD-40-C-AC.

The EOM drive signal is created using the Agilent E8241A Signal Generator (SG), labelled EOM SG in the paper, and has a range of 250 kHz to 20 GHz. The -20 dB coupler used to deliver the EOM drive signal is Minicircuits ZCDC-01203-S+. The amplifiers for the EOM drive signal are the 15 dB Marki Microwave A-1844 and the 32 dB Advanced Microwave WPA263PA while the PIN modulator is HP 33001C. The frequency doubler is Marki MLD1640LS and the amplifier after the doubler is the 17 dB Marki A-2050. The mixer used is Marki MM1-1857L. 

The downconverted beat note is amplified by 83.5dB using three amplifiers in the following order: MiniCircuits ZFL-500LN-BNC+ (24.5 dB), ZFL-1200GH+ (34 dB), ZFL-2500VH+ (25 dB). The beat note is then passed through two MiniCircuits BBP-101+ bandpass filters. The -10 dB coupler used is MiniCircuits ZFDC-10-1 and the power splitter is MiniCircuits ZSC-2-1+. The phase locked loop (PLL) reference signal is created with the HP 8665A Signal Generator, which has a range of 100 kHz to 4.2 GHz and is labelled as PLL SG in the paper. The frequency counter used is Berkeley Nucleonics 1105. 

The dual comparator used in the PLL circuit is AD96687BQ \cite{AD96687}. The 4-bit counter used is MC10H016P \cite{MC10H016}. The phase and frequency discriminator used is AD9901 \cite{AD9901}. The instrumentation amplifier used in the error generator is AD620 while the op-amp used in the loop filter is LM324.


\section{Additional Results}

\begin{figure}[H]
    \centering
    \includegraphics[width = \textwidth]{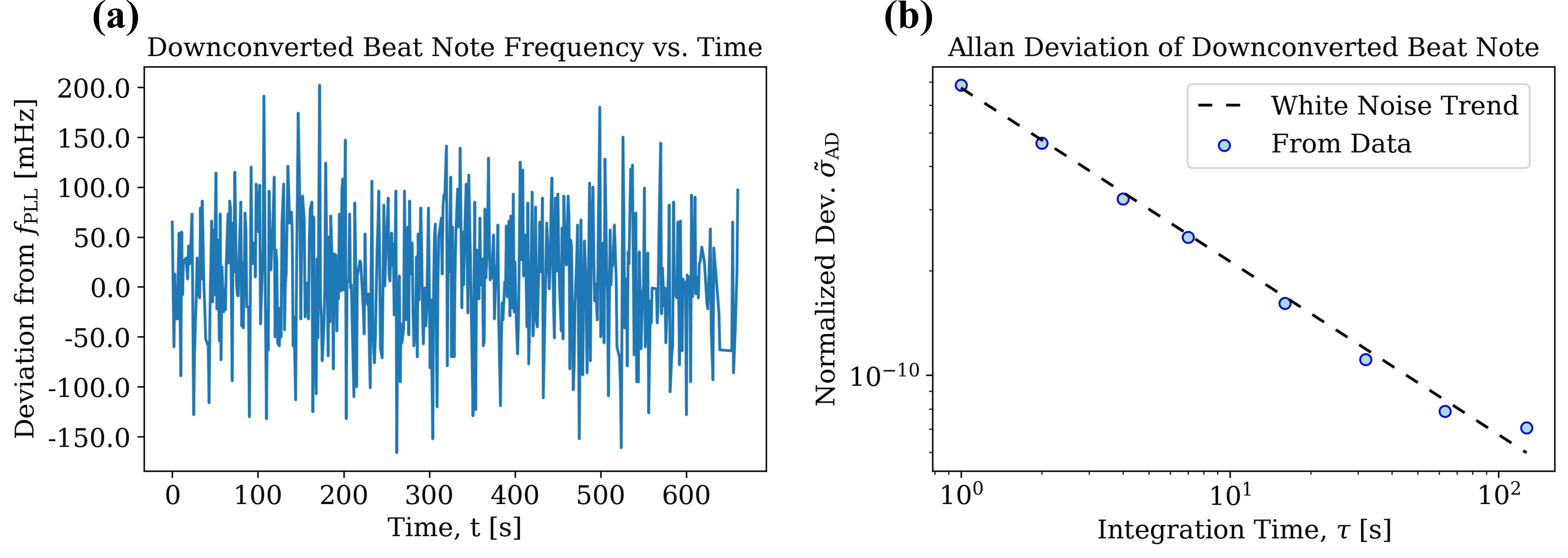}
    \caption{\textbf{(a).} A time signal of the processed beat note frequency over 10 minutes. We set $f_{\mathrm{PLL}} = 100$ MHz. Recorded at an offset frequency of +220.1 GHz. \textbf{(b).} The corresponding normalized Allan deviations calculated for integration times from 1 to 128 s. The normalized deviations $\Tilde{\sigma}_{\mathrm{AD}}$ were found by dividing the Allan deviations by the mean frequency of the time signal. The dashed line represents the Allan deviation behavior for white noise, where $\Tilde{\sigma}_{\mathrm{AD}} \propto \tau^{-1/2}$}
    \label{fig: timesigAD}
\end{figure}

\tab The PLL locks the frequency of the processed beat note to the PLL reference signal frequency $f_{\mathrm{PLL}}$, which we set to 100 MHz. To quantify the stability of the offset frequency, the processed beat note frequency is measured by a frequency counter. The time behavior of the beat note frequency over approximately 10 minutes for a typical measurement is shown in Fig. \ref{fig: timesigAD}(a). From the frequency data, the normalized Allan deviations $\Tilde{\sigma}_{\mathrm{AD}}$ were calculated for integration times, $\tau$, from 1 to 128 s and plotted in Fig. \ref{fig: timesigAD}(b). The good agreement between the data and white noise trend line indicates the frequency noise is likely dominated by white noise. 

For each desired offset frequency $\Delta f_{\mathrm{des}}$, the mean and the standard deviation, $\sigma_{\Delta f}$, of the processed beat note frequency is found from the recorded time signal. The mean offset frequency $\overline{\Delta f}$ is found by adding $nf_{\mathrm{EO}} + f_{\mathrm{LO}} = n(20$  GHz) + 40 GHz to the mean beat note frequency frequency, where $n$ is the corresponding sideband order. The offset variance Var$({\Delta f}) = (\overline{\Delta f} - \Delta f_{\mathrm{des}})^2$ is then found. The maximum beat note power $P_{\mathrm{max}}$ is also recorded. The results are tabulated in Table \ref{tab: results}. Within the offset range of $\pm$ 220.1 GHz, the standard deviation $\sigma_{\Delta f}$ is around 60 mHz while the offset variance Var$(v_{\Delta f})$ is close to or below 100 (mHz)$^2$. The maximum beat power $P_{\mathrm{max}}$ hovers around 21 dBm for offsets between $+$160.1 GHz and -180.1 GHz. Outside of this range, $P_{\mathrm{max}}$ decreases exponentially. 

\begin{table}[H]
\centering
\caption{\label{tab: results} Calculated quantities for each offset frequency.}  
 \begin{tabular}{ c  c  c  c  c } 
 \hline
$\Delta f_{\mathrm{des}}$ [GHz] & $\sigma_{\Delta f}$ [mHz] & Var$({\Delta f})$ [$\mathrm{(mHz)^2}$] & $P_{\mathrm{max}}$ [dBm] & Sideband Order, $n$ \\
 \hline
 +240.1 & 2.244 $\times 10^8$ & 8.203 $\times 10^{16}$ & -1.2 & +10 \\
 \hline
 +220.1 & 66.994 & 80.153 & 5.2 & +9 \\ 
 \hline 
 +200.1 & 59.958 & 100.658 & 12.8 & +8 \\
 \hline
 +180.1 & 63.582 & 1.940 & 18.0 & +7 \\
 \hline
 +160.1 & 62.260 & 26.172 & 21.2 & +6 \\
 \hline
 +140.1 & 58.485 & 5.170 & 21.2 & +5 \\
 \hline
 +120.1 & 60.635 & 46.528 & 20.8 & +4 \\
 \hline
 +100.1 & 60.611 & 49.282 & 21.6 & +3 \\
 \hline
 +80.1 & 59.891 & 22.259 & 22.4 & +2\\
 \hline
 +60.1 & 59.024 & 24.598 & 22.0 & +1\\
 \hline
 +40.1 & 55.156 & 106.882 & 22.4 & 0\\
 \hline 
 -40.1 & 60.986 & 28.248 & 22.4 & 0\\
 \hline
  -60.1 & 65.248 & 19.785 & 22.0 & -1\\
 \hline
  -80.1 & 60.923 & 42.177 & 22.4 & -2\\
 \hline
  -100.1 &  61.326 & 6.689 & 21.6 & -3 \\
 \hline
  -120.1 & 58.523 & 17.218 & 21.6 & -4 \\
 \hline
  -140.1 & 61.018 & 47.699 & 22.0 & -5 \\
 \hline
  -160.1 & 60.990 & 21.726 & 22.0 & -6 \\
 \hline
  -180.1 & 63.490 & 62.879 & 20.8 & -7 \\
 \hline
 -200.1 & 58.092 & 0.116 & 15.4 & -8 \\
 \hline
 -220.1 & 60.237 & 96.148 & 9.4 & -9 \\ 
 \hline 
 -240.1 & 5.537 $\times 10^4$ & 8.897 $\times 10^9$ & 2.8 &  -10 \\
 \hline 
\end{tabular}
\end{table}

At higher order sidebands, there is an anisotropy in $P_{\mathrm{max}}$ for blue versus red offset frequencies, which is due to residual amplitude modulation (RAM) in the EOM \cite{RAM1, RAM2, RAM3, RAM4, RAM5}. RAM is caused by a combination of factors including multiple reflections off of the EOM crystal face \cite{RAM1}, EOM crystal temperature variations \cite{RAM3}, the photorefractive effect \cite{RAM3}, and spatial inhomogeneities of the electric field in the EOM crystal \cite{RAM1}.

\section{Calculating Loop Bandwidth}
\tab Our loop filter outputs a control signal $v_\mathrm{c}(t) = K_p e(t) + K_i\int e(t) dt$ where $K_p = \frac{R_2}{R_1}$ is the proportional gain and $K_i = \frac{1}{R_1 C}$ is the integral gain \cite{PLLBook1, PLLBook2}. For our loop filter, $C$ = 0.109 $\mu$F and we found the optimal resistance values to be $R_1 = 460$ $\Omega$ and $R_2 = 687$ $\Omega$, corresponding to $K_p = 1.49$ and $K_i = 19944.16$ $\mathrm{s}^{-1}$. We use the unity gain crossover frequency definition of loop bandwidth for a PLL with an active PI loop filter \cite{PLLBook1, PLLBook2}. Using the loop filter gains $K_i, K_p$, the frequency division factor $N$, the controlled laser voltage sensitivity $K_o$, and the PFD gain $K_d$, we calculate the loop bandwidth to be $f_{\mathrm{BW}}$ = 50 kHz, which is of the same order of magnitude as the measured bandwidth of 24 kHz.

\bibliography{sample}
